\documentclass[cameraready]{Interspeech}

\title{HRTF-guided Binaural Target Speaker Extraction with Real-World Validation}

\author[affiliation={1}, orcid=0009-0001-6116-2869]{Yoav}{Ellinson}
\author[affiliation={1}, orcid=0000-0002-2885-170X]{Sharon}{Gannot}

\address{
    $^1$ Faculty of Engineering, Bar Ilan University, Israel
}

\email{yoav.ellinson@biu.ac.il, sharon.gannot@biu.ac.il}

\keywords{Binaural audio,Target Speaker Extraction, HRTF}

\usepackage{comment}
\usepackage{acronym}
\usepackage{tikz}
\usetikzlibrary{positioning,arrows.meta,fit,calc,shapes.geometric}
\usepackage{subcaption,balance}
\usepackage{multirow}
\usepackage{mathtools}

\acrodef{HRTF}{Head-Related Transfer Function}
\acrodef{HRIR}{Head-Related Impulse Response}
\acrodef{TrF}{Transfer Function}
\acrodef{TSE}{Target Speaker Extraction}
\acrodef{TF}{Time-Frequency}
\acrodef{RIR}{Room Impulse Response}
\acrodef{BRIR}{Binaural Room Impulse Response}
\acrodef{STFT}{Short-Time Fourier transform}
\acrodef{ReLU}{Rectified Linear Unit}
\acrodef{DOA}{Direction of Arrival}
\acrodef{SI-SDR}{Scale Invariant Signal to Distortion Ratio }
\acrodef{FFT}{Fast Fourier Transform}
\acrodef{BSS}{Blind Source Separation}
\acrodef{MAE}{Mean Absolute Error}
\acrodef{WSJ}{Wall Street Journal}
\acrodef{SOFA}{Spatially Oriented Format for Acoustics}
\acrodef{SR}{Sampling Rate}
\acrodef{IR}{Impulse Response}
\acrodef{PESQ}{Perceptual Evaluation of Speech Quality}
\acrodef{MOS}{Mean Opinion Score}
\acrodef{ILD}{Interaural Level Difference}
\acrodef{ITD}{Interaural Time Difference}
\acrodef{STOI}{Short-time Objective Intelligibility}
\acrodef{RTF}{Relative Transfer Function}
\acrodef{DNN}{Deep Neural Network}
\acrodef{SIR}{Signal-to-Interference Ratio}
\acrodef{p.d.f.}{Probability Density Function}
\acrodef{LCMV}{Linearly Constrained Minimum Variance}
\acrodef{BLCMV}{Binaural Linearly Constrained Minimum Variance}
\acrodef{RIR}{Room Impulse Response}
\acrodef{LS}{LibriSpeech}
\acrodef{MLS}{Multilingual LibriSpeech}
\acrodef{MIMO}{Multi input Multi output}
\acrodef{Bi-TSE}{Binaural Target Speaker Extraction}
\acrodef{Bi-TSE-HRTF}{Binaural Target Speaker Extraction using HRTFs}
\acrodef{RI}{Real-Imaginary}
\acrodef{MOS}{Mean Opinion Score}
\acrodef{NBSS}{Narrow-band Deep Speech Separation}
\acrodef{HATS}{head and torso simulator}
\acrodef{NISQA}{Non-Intrusive Speech Quality Assessment}
\begin{document}
\maketitle
\begin{abstract}
This paper presents a \ac{HRTF}-guided framework for binaural \ac{TSE} from mixtures of concurrent sources. Unlike conventional \ac{TSE} methods based on \ac{DOA} estimation or enrollment signals, which often distort perceived spatial location, the proposed approach leverages the listener’s \ac{HRTF} as an explicit spatial prior.
The proposed framework is built upon a multi-channel deep blind source separation backbone, adapted to the binaural \ac{TSE} setting.
It is trained on measured \acp{HRTF} from a diverse population, enabling cross-listener generalization rather than subject-specific tuning. By conditioning the extraction on \ac{HRTF}-derived spatial information, the method preserves binaural cues while enhancing speech quality and intelligibility.
The performance of the proposed framework is validated through simulations and real recordings obtained from a \ac{HATS} simulator in a reverberant environment.

\end{abstract}

\section{Introduction}

Humans possess a remarkable ability to selectively attend to a desired sound source even in acoustically adverse environments characterized by background noise, reverberation, and multiple concurrent speakers. This phenomenon, referred to as the cocktail party problem~\cite{cherry1953}, arises when several sources and environmental distortions are captured simultaneously.
Such conditions are particularly challenging for hearing-impaired listeners, whose selective auditory attention is severely degraded.

The \ac{TSE} paradigm provides a principled framework for addressing this problem. In contrast to conventional \ac{BSS}, which aims to separate all active sources without prior knowledge, \ac{TSE} assumes the availability of auxiliary information about the target speaker, thereby enabling directed extraction of the desired signal.
Recently, there has been significant interest in leveraging deep learning models to address this challenge. A comprehensive overview of the task and recent advances is provided in~\cite{zmolikova2023neural}. Studies indicate that the auxiliary information used for extraction, commonly referred to as the extraction cue, may be visual \cite{lin2023av,pan2022usev,pan2023imaginenet}, spectral \cite{eisenberg2022single,delcroix2018single}, spatial \cite{eisenberg2025end,10096098}, or a combination thereof.

This work addresses the \ac{TSE} problem in a binaural audio setting, where no visual modality is available, and the acoustic scene is captured by two microphones corresponding to the listener’s left and right ears. This configuration inherently provides spatial information, as the binaural signals encode interaural time and level differences that facilitate source localization and spatial segregation. Consequently, the two-microphone setup offers a natural framework for exploiting spatial cues in target extraction.
In contrast, approaches that rely primarily on spectral characteristics—such as those employing an enrollment utterance of the desired speaker—may suffer performance degradation when interfering speakers exhibit similar vocal traits~\cite{delcroix2020improving}. In such cases, limited discriminability in the spectral domain can lead to leakage or target distortion, highlighting the importance of incorporating robust spatial information into the extraction process.

In the context of binaural \ac{TSE}, several recent approaches have been proposed. Methods such as \cite{wang2025leveraging} employ the \ac{DOA} of the desired source as an explicit extraction cue, guiding the model toward signals arriving from a specific direction. In contrast, \cite{meng2024binaural} utilizes a binaural enrollment signal and aims to distill spatial attributes from it in addition to speaker-dependent spectral characteristics.
More recently, \cite{ellinson2026binauraltargetspeakerextraction} proposed leveraging a specific listener’s \ac{HRTF} as the conditioning signal. In this formulation, the raw \ac{HRTF} corresponding to the desired source direction relative to the listener’s head is provided to the model as the extraction cue. By conditioning on the individualized acoustical properties encapsulated in the \ac{HRTF}, the resulting binaural signal is spatially anchored to the intended location, thereby preserving the perceived source location.

Preserving spatial information is a fundamental requirement in binaural audio processing, as spatial cues contribute not only to localization but also to speech intelligibility and perceptual scene organization. The primary binaural cues—\ac{ILD} and \ac{ITD}—encode direction-dependent characteristics of a sound source. 
Maintaining these cues in the extracted signal is therefore essential to ensure spatial consistency and perceptual coherence.
This requirement becomes particularly critical when such algorithms are deployed on wearable devices that directly stream the processed binaural signal to the listener. If spatial cues are distorted, a mismatch may arise between the visually perceived speaker location and the auditory location conveyed by the processed signal. Such inconsistencies between auditory and visual spatial cues can disrupt multisensory integration, impair localization accuracy, and result in an unnatural or cognitively demanding listening experience. In complex acoustic scenes, as originally emphasized in the context of the cocktail party problem~\cite{cherry1953}, spatial alignment plays a central role in effective auditory attention and source segregation.

In this work, we propose a binaural \ac{TSE} framework whose parameters are not tailored to a specific listener. The model is trained on a diverse collection of measured \acp{HRTF}, enabling it to generalize across listeners. 
At inference time, the extraction is conditioned on the \ac{HRTF} associated with the target source's direction. In this manner, the \ac{HRTF} serves as an explicit spatial prior that guides the model toward the intended source while preserving binaural consistency.

\section{Problem Formulation}
This work focuses on scenarios involving multiple concurrent speakers in a binaural setting. Throughout this paper, the superscript $B$ denotes binaural signals and systems, represented as two-channel vectors, e.g., $\mathbf{x}^B = [x_L, x_R]^\top$, where $x_L$ and $x_R$ correspond to the left- and right-ear channels, respectively.
The specific problem variant addressed in this work considers two concurrent speakers in a reverberant environment.

\subsection{Formulation and Notations}
Let $\mathbf{x}^B(n)$ denote the dual-channel (binaural) speech mixture signal, with $n$ denoting the discrete-time index. Let $\mathbf{x}^B(k,\ell)$ denote the \ac{STFT} representation of $\mathbf{x}^B(n)$, where $k$ and $\ell$ are the frequency-bin and time-frame indexes, respectively. Denote $K$, the total number of frequency bins, and $L$, the total number of time frames.
Recall that the mixed signal consists of the sum of the active sources, each convolved with its corresponding \ac{BRIR}. Accordingly, define:
\begin{equation}
    \mathbf{x}^B(k,\ell) = y_1(k,\ell)\mathbf{h}^B_1(k) + y_2(k,\ell)\mathbf{h}^B_2(k)
    \label{eq:signal}
\end{equation}
where $y_s(k,\ell)$ denotes the speech signal of the individual speaker $s\in\{1,2\}$ in the \ac{STFT} domain, and $\mathbf{h}^B_s(k)$ denotes the corresponding \acp{BRIR} associated with each speaker location relative to the sensors. The location of source $s$ is specified in spherical coordinates by its azimuth, elevation, and radial distance, denoted by $\theta_s$, $\phi_s$, and $r_s$, respectively.

In reverberant environments, acoustic propagation between a sound source and a receiver is governed not only by the direct path but also by multiple reflections from surrounding surfaces. To define the resulting \ac{BRIR}, we first introduce $\{\mathbf{h}^{\mathrm{hrtf}}(\theta,\phi,n)\}_{\theta,\phi}$, which denotes a collection of \acp{IR} measured for each azimuth elevation pair on a spherical surface, with the subject's head positioned at the center, in an anechoic environment. Its frequency domain counterpart is denoted by $\{\mathbf{h}^{\mathrm{hrtf}}(\theta,\phi,k)\}_{\theta,\phi}$. Since most \ac{HRTF} databases employ a constant measurement radius, each \ac{HRTF} can be uniquely identified by its azimuth $\theta$ and elevation $\phi$. 

Following the image source method \cite{allen1979image}, the \ac{BRIR} for source $s$ is modeled as a sum of delayed and attenuated \acp{HRTF}:
\begin{equation}
    \mathbf{h}^B_s(n) = \sum_{m=0}^{M} \alpha_{s,m} \mathbf{h}^{\mathrm{hrtf}}(\theta_{s,m},\phi_{s,m},n-\tau_{s,m}),
    \label{eq:brir}
\end{equation}
where $M$ denotes the reflection order. Here, $m=0$ represents the direct propagation path, while $m > 0$ represents subsequent reflections. $\alpha_{s,m}$ is a scalar gain representing the attenuation associated with acoustic propagation (including sound absorption from the room facets), and $\tau_{s,m}$ denotes the corresponding reflection delay measured in samples. These parameters are functions of the source location and are applied identically to both channels, as the \ac{HRTF} inherently accounts for interaural level and time differences. In free-field conditions, the direct-path gain follows the inverse distance law, $\alpha_{s,0} \sim \frac{1}{r_s}$, and is further attenuated due to absorption by reflective surfaces. The delay $\tau_{s,m}$ is determined by the propagation distance of the corresponding reflection path.
The time-domain \acp{BRIR}, $\mathbf{h}^B_s(n)$, admit a frequency-domain representation obtained via the \ac{FFT}, denoted by $\mathbf{h}^B_s(k)$.
Finally, combining \eqref{eq:signal} and \eqref{eq:brir}, the binaural mixture in the \ac{STFT} domain can be expressed as follows:
\begin{multline}
\mathbf{x}^B(k,\ell) = \\ 
\sum_{s=1}^{2} y_s(k,\ell) \sum_{m=0}^{M} \alpha_{s,m} \mathbf{h}^{\mathrm{hrtf}}(\theta_{s,m},\phi_{s,m},k) e^{\frac{-j2\pi k \tau_{s,m} }{K}}. \label{eq:signal_full}
\end{multline}

\subsection{Task}
Our objective is to extract a desired target speaker from the binaural mixture signal $\mathbf{x}^B(n)$, assuming that the spatial location of the target speaker is known. In addition to source extraction, we aim to perform dereverberation. Accordingly, we define the target output signal $\tilde{\mathbf{y}}^B_s(n)$ as binaural signal consisting of the desired speaker $y_s(n)$ convolved with the direct-path \ac{BRIR} corresponding to speaker $s$, obtained by setting $M=0$.
The target signal in the \ac{STFT} domain is, therefore, given by:
\begin{equation}
\tilde{\mathbf{y}}^B_s(k,\ell)
    =
    y_s(k,\ell)
    \mathbf{h}^{\mathrm{hrtf}}(\theta_s,\phi_s,k).
    \label{eq:target}
\end{equation}
Note that the gain associated with the direct-path \ac{HRTF} is omitted, as it can be compensated during output signal normalization and does not affect the validity of the formulation. Furthermore, this enables the direct-path \ac{HRTF}, $\mathbf{h}^{\mathrm{hrtf}}(\theta_s,\phi_s,k)$, to serve as the extraction clue for sources located at different radial distances from the listener, provided they share the same azimuth and elevation angles.

Another important aspect of our formulation is the spatial consistency between the input and output signals. Specifically, we aim to preserve the perceived location of the desired speaker in the extracted output by maintaining the associated binaural spatial cues. The proposed formulation explicitly facilitates this preservation.
The precedence effect \cite{haas1972influence} states that, although the \ac{BRIR} comprises multiple reflections, the direct path is perceptually dominant. Accordingly, we adopt the direct-path \ac{HRTF} as the extraction clue, as it is closely aligned with the problem formulation.

\begin{figure*}[ht!]
    \addtolength{\belowcaptionskip}{-15pt}
    \addtolength{\abovecaptionskip}{-6pt}
    \centering
    \scalebox{0.8}{\definecolor{lilacfill}{RGB}{210, 190, 255}
\definecolor{lilacborder}{RGB}{130, 90, 200}

\begin{tikzpicture}[
    font=\small,
    >=Latex,
    node distance=3mm and 3mm,
    box/.style={
        draw,
        rounded corners=2mm,
        line width=0.8pt,
        align=center,
        minimum height=10mm,
        minimum width=30mm,
        inner sep=4pt
    },
    inp/.style={
        box,
        fill=cyan!22,
        draw=cyan!65!black
    },
    pre/.style={
        box,
        fill=yellow!28,
        draw=yellow!60!black
    },
    hrtf/.style={
        box,
        fill=lilacfill,
        draw=lilacborder!80!black
    },
    core/.style={
        box,
        fill=red!22,
        draw=red!70!black,
        line width=0.9pt
    },
    output/.style={
        box,
        fill=purple!25,
        draw=purple!65!black
    },
    dashedbox/.style={
        draw,
        rounded corners=2mm,
        line width=0.8pt,
        dashed,
        inner sep=6pt
    },
    arrow/.style={->, thick},
    encoder/.style={
        draw,
        trapezium,
        trapezium left angle=65,
        trapezium right angle=65,
        shape border rotate=180,
        thick,
        align=center,
        minimum height=10mm,
        minimum width=10mm,
        shading=axis,
        top color=teal!10,
        bottom color=teal!50
    },
    decoder/.style={
        draw,
        trapezium,
        trapezium left angle=65,
        trapezium right angle=65,
        shape border rotate=90,
        line width=0.8pt,
        align=center,
        minimum height=10mm,
        minimum width=10mm,
        shading=axis,
        top color=teal!10,
        bottom color=teal!50
    },
    every node/.style={font=\footnotesize}
]

\node[inp] (mix) {\shortstack{
Binaural Mixture \\ STFT\\
$\bm{x}^B\in\mathbb{C}^{B\times 2\times K\times L}$
}};

\node[pre,below=of mix] (norm) {\shortstack{
Norm + Re-shape \\ $\bm{x}^B\in\mathbb{R}^{B\cdot K \times L\times 4}$
}};

\node[hrtf, right=of mix, yshift=0.045cm] (h) {\shortstack{
Direct-path\\
HRTF\\
$\bm{h}^{\mathrm{hrtf}}(\theta_s,\phi_s,k)\in\mathbb{C}^{B\times 2\times K}$}};

\node[pre,below = of h] (norm_h) {\shortstack{
Norm +Re-shape \\ $\bm{h}_s^{\mathrm{hrtf}}\in\mathbb{R}^{B \times K \times L\times 4}$
}};

\node[encoder, below=of norm] (enc) {\shortstack{
\textbf{Encoder}:\\
Conv1D \\ $(2C \rightarrow d)$
}};

\node[encoder, below=of norm_h] (enc_hrtf) {\shortstack{
\textbf{Encoder}:\\
Conv2D \\ $(2C \rightarrow d)$
}};

\node[core, right=of norm_h,xshift=0.7cm] (nbc2) {\shortstack{
NBC2 Block}};

\node[decoder, right=of nbc2,xshift=0.3cm] (dec) {\shortstack{
\textbf{Decoder}:\\
Linear \\ $(d \rightarrow 2C)$
}};

\node[output, right=of dec] (y) {\shortstack{
Estimated Target \\ STFT\\
$\hat{\bm{y}}^B_s\in\mathbb{C}^{B\times 2\times K\times L}$
}};

\path let \p1 = (enc), \p2 = (enc_hrtf) in 
  coordinate (mulpos) at ($(0.5*\x1 + 0.5*\x2, -1.4cm+ 0.5*\y1 + 0.5*\y2)$);

\begin{scope}
  \node[draw, circle, minimum size=8mm, thick] (mul) at(mulpos)  {};

  \draw[thick] 
    ($(mul.center) + (-3mm, -3mm)$) -- ($(mul.center) + (3mm, 3mm)$);
  \draw[thick] 
    ($(mul.center) + (-3mm, 3mm)$) -- ($(mul.center) + (3mm, -3mm)$);
\end{scope}

\node[draw, circle, minimum size=7mm, line width=0.8pt] (mul) at(mulpos) {};

\draw[arrow] (mix) -- (norm);
\draw[arrow] (norm) -- (enc);
\draw[arrow] (h) -- (norm_h);
\draw[arrow] (norm_h) -- (enc_hrtf);
\draw[arrow] (enc.south) -- (mul);
\draw[arrow] (enc_hrtf.south) -- (mul);
\draw[arrow] (mul.east)
            --++(3.45cm,0)
            --++(0,2.76cm)
            --(nbc2.west);

\draw[arrow] (nbc2) -- (dec);
\draw[arrow] (dec) -- (y);

\draw[arrow, ->]
    ([xshift=9pt]nbc2.east) 
    -- ++(0,30pt)
    -- ++(-3.7cm,0)
    node[midway, above, font=\footnotesize, text=black!70] {\textbf{Repeat $P$ times}}
    -- ([xshift=-9pt]nbc2.west);


\end{tikzpicture}} 
    \caption{Overview of the proposed binaural target speaker extraction model ($B$ denotes batch size). Based on the \ac{NBSS} framework, the architecture processes the \ac{STFT} mixture by conditioning encoded features on direct-path \acp{HRTF} via latent-space modulation. A stack of $P=8$ NBC2 blocks produces the final complex-valued spectral estimate. See \cite[Fig. 2]{quan2022nbc2} for a detailed NBC2 block illustration.}
        \label{fig:full_arc}
\end{figure*}
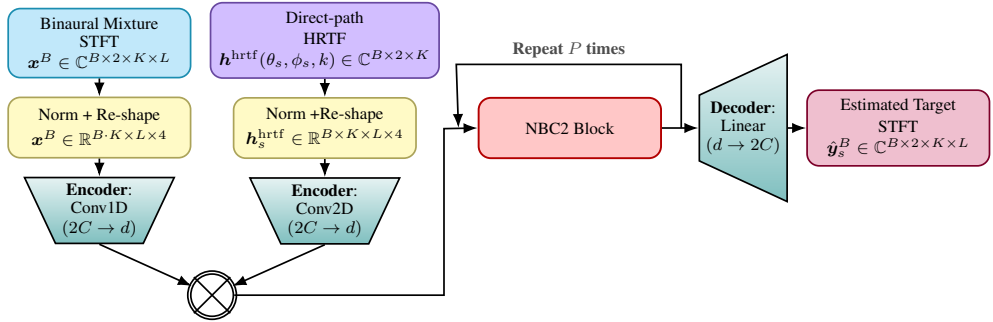

\section{Method}
Following the problem formulation, we aim to extract $\tilde{\mathbf{y}}^B_s(n)$ given $\mathbf{x}^B(n)$, and the desired source location $(\theta_s,\phi_s,r_s)$. Denote the estimated binaural signal as $\hat{\mathbf{y}}^B_s(n)$.
Under the far-field assumption, the radial distance $r_s$ can be absorbed into the propagation scale factor $\alpha_{s,0}$ and omitted from the spatial clue, as \acp{HRTF} primarily represent directional filtering.

\subsection{Model Architecture}
\label{model_arc}
The model architecture employed in this work is based on the \ac{NBSS} framework \cite{quan2022multi}. The model operates in the \ac{STFT} domain and processes binaural mixtures on a per-frequency basis.
The model input consists of the binaural mixture $\mathbf{x}^B(k,\ell)$ and the target spatial clue $\mathbf{h}^{\mathrm{hrtf}}(\theta_s,\phi_s,k)$. Both complex-valued inputs are first transformed by concatenating their real and imaginary parts. These representations are then processed by separate convolutional encoders to project them into a shared latent space.
To incorporate spatial information, the encoded \ac{HRTF} features are replicated along the time dimension and used to modulate the mixture representation through element-wise multiplication in the latent space. This operation conditions the separation process on the \ac{HRTF} corresponding to the target speaker location. The conditioned features are subsequently processed by a stack of NBC2 self-attention blocks, which are designed to capture correlations within frequency bands. When conditioned on the \ac{HRTF}, these blocks emphasize spectral components that are consistent with the desired spatial configuration, thereby extracting the speech corresponding to the target source location encoded by the provided \ac{HRTF}.
Finally, a linear decoder maps the latent features to complex-valued spectral estimates, which are rescaled and returned in the \ac{STFT} domain. Fig.~\ref{fig:full_arc} illustrates the complete architecture.

\subsection{Loss-Function}
We used the \ac{SI-SDR}~\cite{le2019sdr} as the primary loss function during training by computing the mean \ac{SI-SDR} across both channels, as in \eqref{eq:sisdr}. 
\begin{multline}  \mathrm{SI\text{-}SDR}^B(\tilde{\mathbf{y}}^B_s,\hat{\mathbf{y}}^B_s)
    = \\
    \frac{1}{2}\big(\text{SI-SDR}(\tilde{y}_{s,L},\hat{y}_{s,L})+ \text{SI-SDR}(\tilde{y}_{s,R},\hat{y}_{s,R})  \big)
    \label{eq:sisdr}
\end{multline}
Additionally, we employed an \ac{MAE} loss in the \ac{STFT} domain, presented in \eqref{eq:mae}. 
\begin{multline}
    \mathrm{MAE}(\tilde{\mathbf{y}}^B_s(k,\ell),\hat{\mathbf{y}}^B_s(k,\ell))
    =  \\
    \frac{1}{K L}
    \sum_{k=0}^{K-1}
    \sum_{\ell=0}^{L-1}
    \lVert
    \tilde{\mathbf{y}}^B_s(k,\ell) - \hat{\mathbf{y}}^B_s(k,\ell)
    \rVert_1
    \label{eq:mae}
\end{multline}
Both loss terms were employed throughout most of the training process, whereas in the final epochs only the \ac{SI-SDR} loss was used for fine-tuning.
\section{Experimental Setup}
\label{sec:experimental_setup}
We evaluate the proposed HRTF model using a combination of large-scale simulated reverberant mixtures and real-world recordings captured with a \acf{HATS}.
\subsection{Data Simulation}
All models were trained and evaluated using the WSJ0 speech corpus \cite{garofolo2007csr}. Binaural signals were simulated using the SofaMyRoom framework \cite{barumerli2021sofamyroom} to generate reverberant \acp{BRIR}. The reverberation time was drawn from $T_{60}\sim\mathcal{U}[0.2, 0.8]$~s, and fully overlapped speech mixtures were created with an \ac{SIR} drawn from $\mathcal{U}[-5, 5],$dB.
Individualized \acp{HRTF} measurements were provided in the \ac{SOFA} format \cite{majdak2013spatially,majdak2022spatially} and obtained from the \ac{SOFA} conventions repository \cite{sofa_conventions}.
Only measured \ac{HRTF} data were used in this study, drawn from the following datasets: ARI \cite{ari_hrtf}, SONICOM \cite{engel2023the}, RIEC \cite{watanabe2014dataset}, SADIE \cite{armstrong2018perceptual}, SS2 \cite{warneckeHRTF2024}, Viking \cite{spagnol2020viking}, and HRIR CIRC360 \cite{bernschutz2020spherical}.
In total, 789 measured \acp{HRTF} from distinct subjects were used for training and validation, while 7 unseen subjects, one from each dataset, were reserved for testing.
Each simulated mixture comprises two speakers and can therefore be used for target speaker extraction in two symmetric configurations. During training and validation, the target speaker was selected at random, allowing each mixture to contribute two distinct training examples over epochs. During testing, extraction was performed for both speakers in each mixture, resulting in effective dataset sizes of 16k, 4k, and 2k utterances for training, validation, and testing, respectively.

\subsection{Implementation}
Audio signals were sampled at 16~kHz, cut or zero padded to 5~s long and transformed to the \ac{STFT} domain using a 512-point window with 75\% overlap (257 bins). Following the \ac{NBSS} framework, we used $P=8$ NBC2-small blocks as specified in \cite[Sec.~V-B]{quan2022nbc2}. Training employed the AdamW optimizer for 260 epochs at a $10^{-3}$ learning rate, followed by 30 fine-tuning epochs at $10^{-4}$. During fine-tuning, the \ac{MAE} loss was disabled to maximize \ac{SI-SDR}.

\subsection{Competing Method}
A natural competing approach to the proposed method is \cite{wang2025leveraging}, which employs the same \ac{NBSS}-small backbone but relies on a different extraction clue. We trained this model using the exact same training dataset and strictly followed the configuration settings prescribed by the authors. Specifically, we adopted the leading configuration reported in their work, denoted as BDE+CDF+IPD+SDF, and refer to it hereafter as DOA-BDE.

\subsection{Real-World Recordings}
The models were further evaluated using real binaural recordings captured in a reverberant room with $T_{60}=0.37$~s. These recordings were conducted using a \ac{HATS} (Brüel \& Kjær Type~4128C) mounted on a precision turntable (Brüel \& Kjær Type~9640) providing an angular resolution of $1^\circ$. The \ac{HATS} was surrounded by a quarter-circular loudspeaker array with a radius of $r\approx1.5$~m and loudspeaker heights ranging from $-30$~cm to $+30$~cm relative to the \ac{HATS} ear level. All source heights and radii were measured precisely to enable an exact calculation of the elevation angles from the source to the receiver.
This configuration enabled recordings across a wide range of azimuth and elevation combinations, as well as controlled angular distance (azimuth) between concurrent speakers, ranging from $20^\circ$ to $90^\circ$ in $10^\circ$ increments. For each of the eight angular distances, 30 mixtures were recorded, resulting in a total of 240 samples.
Both the proposed method and the competing method were applied to the captured mixtures. The proposed approach was conditioned on \acp{HRTF} extracted from the \ac{HATS} database, which is part of SS2 database 
\cite{warneckeHRTF2024}, whereas the competing method was provided with the ground-truth \ac{DOA} of the desired speaker.

\subsection{Metrics}
Extraction performance is evaluated using the \ac{SI-SDR} improvement (\ac{SI-SDR}i), defined in \eqref{eq:sisdr_i} as the difference between the output and input \ac{SI-SDR} calculated via \eqref{eq:sisdr}. 
\begin{equation}
    \mathrm{SI\text{-}SDRi}
    =
    \mathrm{SI\text{-}SDR^B}(\hat{\mathbf{y}}^B_s,\tilde{\mathbf{y}}^B_s)
    -
    \mathrm{SI\text{-}SDR^B}(\mathbf{x}^B,\tilde{\mathbf{y}}^B_s).
\label{eq:sisdr_i}
\end{equation}
We additionally employ the \ac{PESQ} metric \cite{rix2001perceptual} to assess perceptual quality.
The former reflects the correlation between the estimated and target signals, whereas the latter assesses perceived speech quality.
For real recordings, where target signals are unavailable, we used a non-intrusive metric to assess the algorithm's performance. Thus, we opted to use \ac{NISQA} \cite{mittag2021nisqa}, a \ac{DNN}-based \ac{MOS} predictor. We report the average \ac{NISQA} score of the left and right channels.

In addition to perceptual quality measures, we assess spatial consistency by comparing the binaural cues of the extracted and target signals. Specifically, we compute histograms of the \ac{ITD} and \ac{ILD} for both signals and measure the deviation between their dominant peaks. These histograms are computed using the procedure in \cite{faller2004source}. Under our formulation, the direct-path component of the \ac{BRIR} is expected to dominate each cue. We denote these deviations by $\Delta \mathrm{ILD}$ and $\Delta \mathrm{ITD}$, measured in dB and~ms, respectively.

\begin{table}[t]
\centering
\addtolength{\abovecaptionskip}{-6pt}
\caption{Extraction performance in terms of \ac{SI-SDR}$_\mathrm{i}$~[dB], \ac{PESQ}, and spatial consistency metrics $\Delta$ITD~[ms] and $\Delta$ILD~[dB]. Arrows indicate whether higher or lower values are better. Reported values correspond to mean scores computed over a test set of 1000 mixtures. Since extraction is performed for both speakers in each mixture, the effective number of evaluated samples is 2000.}
\label{tab:results_sim}
\setlength{\tabcolsep}{2pt}
\small
\begin{tabular}{lcccc}
\toprule
Method & \ac{SI-SDR}$_\mathrm{i}$$\uparrow$ & \ac{PESQ}$\uparrow$ & $\Delta$ITD$\downarrow$ & $\Delta$ILD$\downarrow$ \\
\midrule
Mixture & -- & 1.18 & 1.464 & 0.417 \\
DOA-BDE \cite{wang2025leveraging} & 13.881 & 2.74 & 0.982 & 0.479 \\
Proposed & \textbf{15.770} & \textbf{3.03} & \textbf{0.044} & \textbf{0.349} \\
\bottomrule
\end{tabular}
\vspace{-14pt}
\end{table}

\section{Results and Discussion}
In this section, we present and analyze results from both simulated and real recordings, beginning with the simulated dataset described in Section~\ref{sec:experimental_setup}.
Table~\ref{tab:results_sim} presents the results for both perceptual and spatial consistency metrics. The proposed method demonstrates a clear advantage, outperforming the competing approach across all evaluated metrics. 
While the competing method can extract the target speech, it fails to faithfully reproduce the binaural cues. This spatial degradation likely contributes to lower \ac{SI-SDR} and \ac{PESQ} scores, as these correlation-based metrics are highly sensitive to the phase and temporal misalignments caused by inaccurate spatial reconstruction

Using the \ac{HRTF} as an extraction clue enables the model to selectively retain components consistent with the target spatial configuration. This intrinsic link ensures the reconstructed signal suppresses interference while preserving the binaural cues of the intended source. This mechanism can be interpreted as a learned higher-order ``matched filter,'' in which the model is spatially conditioned to \emph{match} the mixture to the target's unique spatial signature.

Generalization from simulated data to real recordings remains challenging for \ac{DNN}-based algorithms due to distribution shifts and recording mismatches. Nevertheless, both methods successfully extracted the target speech in our real-world experiments, with differences arising in perceptual quality. 
A limitation of the proposed \ac{HRTF}-based approach is that its spatial resolution is inherently limited by the angular sampling density of the available database. In our real-recording setup, the \ac{HATS}-\ac{HRTF} database was measured at $6^\circ$ increments in azimuth and $3^\circ$ in elevation, resulting in a maximum discretization error of $\pm 3^\circ$ and $\pm 1.5^\circ$ in each respective direction. Consequently, conditioning must rely on the nearest available \ac{HRTF} measurement. Despite this discretization, the proposed method remains robust, suggesting nearest-neighbor \ac{HRTF} conditioning is sufficient for high-quality extraction and preserves superior performance even amid angular mismatch.

Table~\ref{tab:results_real} presents the \ac{NISQA} scores for each spatial separation, along with the overall average. The proposed method maintains superior performance even under angular mismatch, indicating greater robustness to spatial inconsistencies.
We further observe that the \ac{NISQA} score increases as the angular distance between concurrent sources grows. This behavior is expected, as smaller separations make it more challenging to distinguish and separate sources using spatial information alone.

Demonstrating robustness on real recordings is further emphasized when considering potential use cases beyond strictly binaural audio settings. The proposed method can be extended to devices equipped with multiple microphones and deployed in real-world scenarios, provided that the relative location of the desired source with respect to the device is available. 

\begin{table}[t]
\centering
\addtolength{\abovecaptionskip}{-6pt}
\caption{Mean \ac{NISQA} scores ($\uparrow$) versus angular distance between speakers, evaluated over 30 real-world recordings per distance, totaling 480 samples (two speakers per recording).}

\label{tab:results_real}

\setlength{\tabcolsep}{1.45pt}
\small

\begin{tabular}{
l
*{9}{c}}
\toprule

Method 
& \multicolumn{1}{c}{$20^\circ$}
& \multicolumn{1}{c}{$30^\circ$}
& \multicolumn{1}{c}{$40^\circ$}
& \multicolumn{1}{c}{$50^\circ$}
& \multicolumn{1}{c}{$60^\circ$}
& \multicolumn{1}{c}{$70^\circ$}
& \multicolumn{1}{c}{$80^\circ$}
& \multicolumn{1}{c}{$90^\circ$}
& \multicolumn{1}{c}{Avg.} \\

\midrule

Mixture &
2.05 & 2.07 & 2.09 & 2.15 & 2.14 & 2.23 & 2.19 & 2.18 & 2.14  \\
\footnotesize{DOA-BDE \cite{wang2025leveraging} }
& 3.00 & 3.03 & 3.05 & 3.17 & 3.16 & 3.25 & 3.26 & 3.20 & 3.14 \\ 
Proposed 
& {\bfseries 3.02} & {\bfseries 3.12} &{\bfseries  3.17} & {\bfseries 3.31} & {\bfseries 3.29} & {\bfseries 3.34} & {\bfseries 3.39} & {\bfseries 3.35} & {\bfseries 3.22} \\ 
\bottomrule
\end{tabular}
\vspace{-14pt}
\end{table}

\section{Conclusions}
We presented a spatially consistent binaural \ac{TSE} framework conditioned directly on the listener's \ac{HRTF}. Unlike prior \ac{DOA}-based approaches, the proposed method incorporates the binaural spatial characteristics into the extraction process. Importantly, the model is trained on a large and diverse collection of measured \acp{HRTF}, enabling generalization across listeners rather than restricting the framework to a single subject.

Experimental results obtained from both simulated data and real-world recordings captured with a \ac{HATS} in a reverberant environment demonstrate that the proposed method preserves binaural spatial cues while achieving significant improvements in perceptual speech quality.
Moreover, evaluation under inherent angular mismatch arising from the finite resolution of the \ac{HRTF} database highlights the model's robustness to localization inaccuracies. These findings suggest that leveraging the listener's \ac{HRTF} as the extraction cue is a practical and effective strategy for achieving spatially consistent \ac{TSE} in real-world scenarios.

\bibliographystyle{IEEEtran}
\bibliography{mybib}

\end{document}